\documentclass[USenglish,oneside,twocolumn]{article}

\usepackage[utf8]{inputenc}
\usepackage[big]{dgruyter_NEW}

\usepackage{epsfig,endnotes}
\usepackage{xspace}
\usepackage{balance}
\usepackage{url}
\usepackage{caption}
\usepackage{subcaption}
\usepackage{color}
\usepackage{hyperref}
\usepackage{balance}
\usepackage{graphicx}
\usepackage{multirow}
\usepackage{xspace}
\RequirePackage{tikz}
\usetikzlibrary{arrows}

\cclogo{\includegraphics{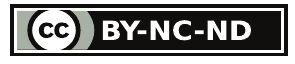}}

\usepackage{pifont} 
 
\newcommand{\cmark}{\ding{51}}%

\newcommand{\system}{ODNS}
\newcommand{\ie}{{\em i.e.\xspace}}
\newcommand{\eg}{{\em e.g.\xspace}}

\graphicspath{ {./figures/} }

\begin{document}
\author*[1]{Paul Schmitt}

  \author[2]{Anne Edmundson}

  \author[3]{Allison Mankin}

  \author[4]{Nick Feamster}

  \affil[1]{Princeton University, E-mail: pschmitt@cs.princeton.edu}

  \affil[2]{Princeton University, E-mail: annie.edmundson@gmail.com}

  \affil[3]{Salesforce, E-mail: allison.mankin@gmail.com}

  \affil[4]{Princeton University, E-mail: feamster@cs.princeton.edu}

  \title{\huge Oblivious DNS: Practical Privacy for DNS Queries}

  \runningtitle{Oblivious DNS: Practical Privacy for DNS Queries}
  
\begin{abstract} 
{Virtually every Internet communication typically involves a Domain Name System (DNS) lookup for the destination server that the client wants to communicate with. Operators of DNS recursive resolvers---the machines that receive a client's query for a domain name and resolve it to a corresponding IP address---can learn significant information about client activity. Past work, for example, indicates that DNS queries reveal information ranging from web browsing activity to the types of devices that a user has in their home. Recognizing the privacy vulnerabilities associated with DNS queries, various third parties have created alternate DNS services that obscure a user's DNS queries from his or her Internet service provider. Yet, these systems merely transfer trust to a different third party. We argue that no single party ought to be able to associate DNS queries with a client IP address that issues those queries. To this end, we present Oblivious DNS (\system{}), which introduces an additional layer of obfuscation between clients and their queries. To do so, \system{} uses its own authoritative namespace; the authoritative servers for the \system{} namespace act as recursive resolvers for the DNS queries that they receive, but they never see the IP addresses for the clients that initiated these queries. We present an initial deployment of \system{}; our experiments show that \system{} introduces minimal performance overhead, both for individual queries and for web page loads. We design \system{} to be compatible with existing DNS protocols and infrastructure, and we are actively working on an open standard with the IETF.}
\end{abstract}
\keywords{privacy, DNS}
  
%
%
\maketitle

\section{Introduction} \label{sec:intro}
Almost all communication on the Internet today starts with a Domain Name System (DNS) lookup. Before communicating with any Internet destination, a user application typically first issues a Domain Name System (DNS) lookup, which takes a domain name (\eg, {\tt google.com}) and returns an IP address for the server that the client should contact. Today, the DNS requires the user to place tremendous trust in DNS operators, who can see all of the DNS queries that a user issues. Whether the operator is an Internet service provider (ISP) or a third party is less concerning than the fact that some single operator can observe and retain this sensitive information. This paper presents a system, called \system{}, which attempts to solve this problem.

As DNS operates today, queries and responses are viewable as plaintext at the recursive resolver, even if the client is using an encrypted channel between it and the recursive resolver. As a result, they can reveal significant information about the Internet destinations that a user or device is communicating with. For example, the domain names themselves reveal the websites that a user visits. Additionally, in the case of smart-home Internet of Things (IoT) devices, DNS queries may reveal the types of devices in user homes. Previous work has also demonstrated that DNS lookups can identify the websites that a user is visiting even when they are using an anonymizing service such as Tor~\cite{greschbach2016effect}. Recursive DNS resolver operators can readily associate and track client identities (\ie, IP addresses) along with information about their DNS queries, creating a fundamental point of privacy risk.

A user's Internet Service Provider (ISP) often operates the user's default recursive
DNS resolver, giving the ISP potentially extraordinary access to DNS query information. To mitigate this risk, several entities, including Google, Cloudflare~\cite{cloudflare_dns}, and Quad9~\cite{quad9} operate ``open'' recursive DNS resolver services that anyone can use as an alternative to their ISP's DNS recursive resolver. Yet, when a user switches to such an alternate resolver, the privacy problem isn't solved; rather, the user must then trust the operator of the open recursive rather than their ISP. Essentially, the user must decide whether they trust their ISP or some other organization---some of which are even in the business of collecting data about users.

Other approaches have layered encryption on top of DNS. For example, DNS-over-TLS~\cite{dns_tls}, DNS-over-DTLS~\cite{dns_dtls}, and DNS-over-HTTPS~\cite{dns_https} send DNS queries over an encrypted channel, which prevents an eavesdropper between the client and the recursive resolver from learning the contents of a DNS lookup but does not prevent the recursive resolver itself from linking queries and IP addresses. DNSCurve uses elliptic curve cryptography to encrypt DNS requests and responses; it also authenticates all DNS responses and eliminates any forged responses~\cite{bernstein2009dnscurve}. DNSCrypt encrypts and authenticates DNS traffic between a client and a recursive resolver~\cite{denis2015dnscrypt}. Neither of these approaches prevent the recursive resolver from observing DNS queries and responses.

This work takes a different tack: Instead of merely shifting the trust anchor from an ISP to some other third party, we seek to prevent a recursive DNS resolver from associating client identities with the queries they make. To do so, we design, implement, and deploy Oblivious DNS (\system{}), which (1)~obfuscates the queries that a recursive resolver sees from the clients that issue DNS queries; and (2)~obfuscates the client's IP address from upper levels of the DNS hierarchy that ultimately resolve the query (\ie, the authoritative servers). \system{} operates in the context of the existing DNS protocol, allowing the existing deployed infrastructure to remain unchanged. \system{} decouples client identity from queries by leveraging the behavior of the global DNS system itself. A client sends an encrypted query to a recursive resolver, which then forwards the query to a \system{} resolver (an authoritative DNS server that can resolve \system{} queries). The recursive resolver never sees the domains that the client queries, and the \system{} resolver never sees the IP address of the client.

\system{} provides benefits for both the recursive resolver operators as well as users of the system. The rich information available at the recursive resolver can make operators the targets of data requests. \system{} reduces the information that operators are able to know as they cannot associate queried domains with client identity, making them less-valuable targets. In essence, the operators are oblivious to their client's requests. Likewise, users of \system{} benefit in that they are no longer required to trust that their recursive resolver has their best interests at heart. Instead, users can be assured that all DNS infrastructure beyond the stub is unable to link their identity with their querying behavior.    

We design \system{} to be immediately deployable alongside existing DNS infrastructure. We implement this functionality in a prototype \system{} stub and \system{} resolver in Go. \system{}'s privacy enhancements come at a performance cost. Our trace-driven evaluation shows that any individual uncached DNS lookup is slower due to two factors: 1)~our cryptographic operations add roughly two milliseconds to the lookup time; and 2)~the round-trip time between the client and the \system{} resolver is added for each lookup. We mitigate the impact of round-trip time latency by designing \system{} to use anycast. Ultimately, our evaluation shows that the performance overhead on web page load times is negligible. Additionally, we reduce the traffic burden placed on existing recursive resolvers by implementing a cache at the stub resolver.
\begin{figure}[t]
    \centering
    \includegraphics[width=.45\textwidth]{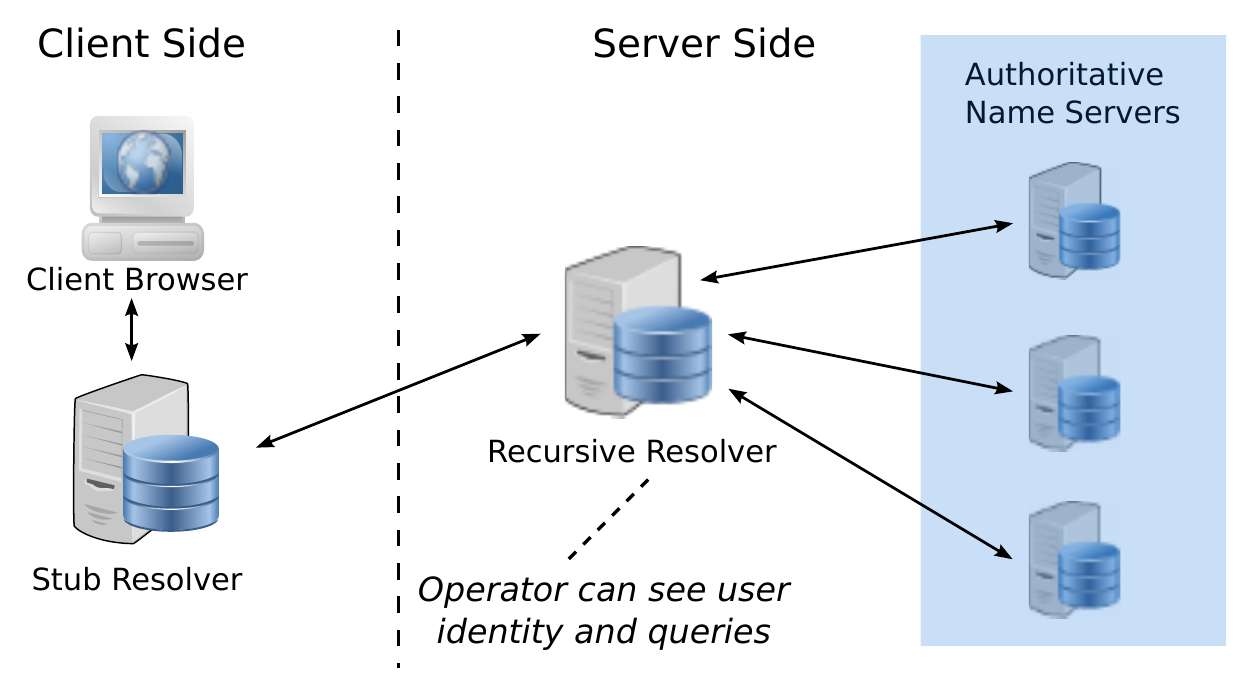}
    \caption{In a typical DNS lookup, a recursive resolver sees DNS queries and responses,
as well as the client IP addresses that issue the queries.}
    \label{fig:dns}
    \vspace{-4mm}
\end{figure}

\section{Background} \label{sec:background}
The domain name system is the hierarchical, decentralized infrastructure responsible for translating between domain names and the associated information that is necessary for data connectivity~\cite{RFC1034, RFC1035}. Clients, such as Internet browsers, issue DNS queries to find the IP address of a server that contains content the user requests. Typically, for a given DNS request, there are up to five DNS servers that can be involved in resolving information for a domain name with each server playing a different role (Figure~\ref{fig:dns}): 1) stub resolver or "client"; 2) recursive resolver; 3) root nameserver; 4) top level domain (TLD) nameserver; and 5) authoritative nameserver.

Stub resolvers are lightweight processes included in operating systems that initiate DNS queries to the larger infrastructure, typically toward a recursive resolver. The recursive resolver will inspect the query and check whether there is a cached answer, if so, the response is returned to the stub. If the answer is not cached, the recursive will then act on the stub's behalf to find the answer. If the uncached query includes a domain for which the recursive server does not already know the address for the domain's authoritative nameserver it will query the root and/or the TLD servers to find the authoritative server IP address. The root server will respond to a query with the appropriate TLD nameserver for that domain (\eg, the {\tt .com} TLD server for a {\tt foo.com} query). The recursive server will query the TLD server, which will in turn respond with the IP address of the authoritative nameserver for the domain. The recursive server will finally query the authoritative nameserver, the final authority for translating a domain name to an IP address. The portion of the DNS namespace that an authoritative server is responsible for is known as a ``zone''. 

\textbf{DNS Privacy Implications.} DNS traffic can reveal personal information about users' browsing behavior as well as the types of devices in a network. The vantage point provided at the recursive DNS resolver gives the operator of the resolver visibility into the IP addresses that query various domain names, which may be ultimately linked to individual devices, sets of devices, or user identity.

Additionally, recursive resolvers can pass information about the user to the entities at higher levels in the DNS hierarchy using the \textit{EDNS0 client subnet} option. EDNS0 client subnet is a non-standards track RFC~\cite{RFC7871}, which is widely-deployed, whereby recursive resolvers include the client IP subnet as part of the DNS query when querying authoritative nameservers. The intent of EDNS0 client subnet is to allow for DNS and content providers to utilize the client information to create more informed responses (\ie, route users to replicas of content they are closest to). The EDNS0 client subnet extension has privacy implications because it reveals a portion of the IP address that issued the initial DNS query to the authoritative server.

\textbf{Why Not Tor?} At first glance, it may appear that DNS privacy can readily be achieved using the Tor~\cite{dingledine2004tor} network. Tor uses layered encryption and a three-hop circuit to provide client anonymity. However, using Tor for DNS privacy comes with several drawbacks. 

\begin{figure}[t]
    \centering
    \includegraphics[width=.37\textwidth]{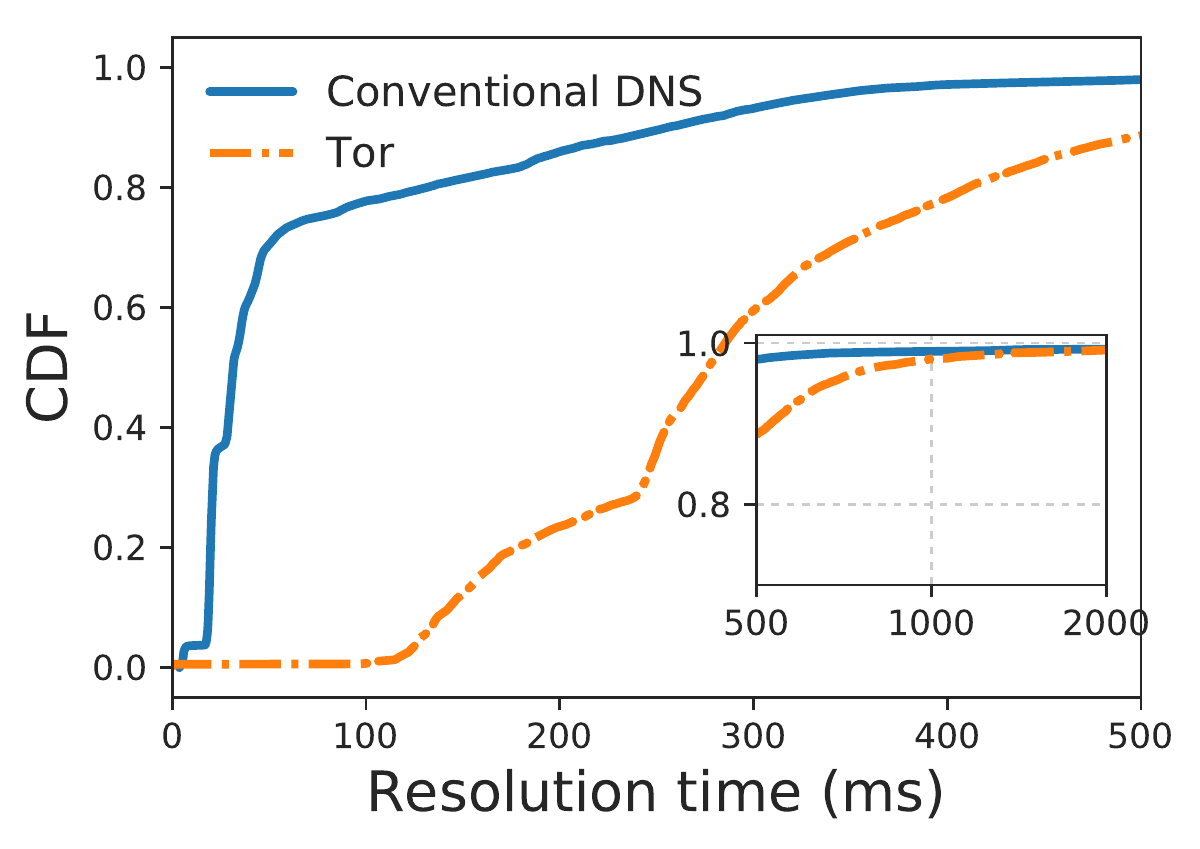}
    \caption{Tor DNS latency compared with conventional DNS.}
    \label{fig:tor}
\end{figure}

First, and perhaps most importantly, Tor's fundamental design introduces substantial network latency to the end-to-end path. This significantly reduces the performance for all traffic over the network. DNS underpins almost all web traffic and increased DNS lookup latency has an outsized impact of web performance~\cite{Sundaresan:2013:CCA:2504730.2504741}. Figure~\ref{fig:tor} provides insight into the additional latency Tor introduces for DNS traffic. We resolve the Alexa top 10,000 domains using both conventional DNS and using the Tor network from a laptop connected to a university network via gigabit Ethernet. As shown, Tor significantly increases the latency for DNS queries. The median resolution time for conventional DNS is 31.31~ms, whereas the median time for DNS via Tor is 276.76~ms.

Next, Tor has long been a target for censorship~\cite{179494}. There are many techniques by which censors can identify Tor nodes and traffic and block them from the Internet. Tor is designed to be all-or-nothing, meaning the client sends all traffic through the circuit. This has both practical impact as censors may block all access to the Internet as well as a performance impact as clients may not require privacy for all of their traffic.    

Finally, while the original client IP may be obfuscated using Tor, the exit node IP address can be associated with all DNS queries that it forwards. Such system behavior leads to a chilling effect as the operators of exit nodes---often individual volunteers---may be held liable for the content that ingresses and egresses onto the Internet from their connection~\cite{vice}. Additionally, DNS in Tor is conducted by the exit node of the circuit and is performed in a manner that depends on how it is configured on that machine. As a result, the operator of that machine (as well as the recursive resolver used by the exit node) can learn what content is being requested. More recently, researchers have analyzed how DNS works in Tor and found that fingerprinting attacks can be performed based on DNS data~\cite{greschbach2016effect}. The Tor Project has also designed and implemented onion services/domains (previously called hidden services), which provide server anonymity; onion domain lookup does not use DNS, but can only be accessed via Tor and suffers from usability issues~\cite{RFC7686}. Recent work has highlighted how onion domain name leakages are a source of privacy leakage as well~\cite{thomas2014measuring}.
\section{Threat Model and Goals}
\label{sec:threat}
In this section, we describe our threat model, outline the capabilities of the
attacker, and introduce the design goals and protections that \system{}
provides.

\textbf{Threat Model. }\label{sec:attacker}Our threat model includes any single non-colluding passive adversary who 
wishes to compromise the confidentiality of a client’s IP {\it and} requested 
domain name.  As an eavesdropper, the adversary has the capability to monitor 
and collect traffic between the following entities: the client resolver and 
the recursive resolver, the recursive resolver and the \system{} resolver, or the 
\system{} resolver and an authoritative server.  Note that \system{} addresses an 
adversary that eavesdrops on {\it one} of these connections, but does not 
address a global passive adversary who can eavesdrop on many or all of these 
connections.  The rest of this section describes the potential threats 
associated with an adversary acting as different components of the \system{} system, 
examples of this adversary in practice, and threats that are not considered 
in this work.

An adversary could impersonate a client of the system, and attempt to perform 
chosen ciphertext attacks by generating queries and inspecting the ciphertext as 
well as the corresponding plaintext.  \system{} prevents this type of adversary from 
learning and associating client IP addresses with their DNS queries.  Our threat 
model does not include an adversary acting as a client (or multiple colluding 
clients) who can send enough DNS requests to impact the availability of the \system{} 
system.

An adversary who acts as a recursive resolver, \system{} resolver, or authoritative 
server has the ability to capture logs of DNS queries and respond to requests for 
data (via subpoena or warrant).  Despite these capabilities, \system{} prevents the 
association of client IP addresses with their corresponding DNS queries.  \system{} 
does not attempt to prevent an adversary such as this from modifying DNS queries 
or dropping DNS queries.
					
The adversary that we consider exists in practice: First, the motivation 
for our system was the rising concern over Internet Service Providers (ISPs) 
operating recursive DNS resolvers and gaining access to user information, as well 
as recent proposals for DNS-over-HTTPS (DoH), which propose to instead send all DNS 
queries from a user’s browser directly to a CDN operator (\eg, Cloudflare), which 
may only exacerbate privacy problems by sending a user’s DNS queries to a different 
party. Second, governments have used the global DNS system to monitor and block 
content. Finally, a government may also request logs from the DNS operator, the 
government could also be colluding with a DNS operator; the DNS operator itself might 
even be an adversary.

This work does not attempt to prevent an adversary from compromising the integrity or 
availability of DNS requests.  We do not consider an adversary who operates both a 
recursive resolver and an authoritative server in an attempt to collude.  

\textbf{Goals. }To defend against the adversary described above, we highlight the design goals for \system{}. Both the recursive DNS resolver and the \system{} resolver have different risks, and therefore requires different protections. Both must be protected by decoupling the client identity (\ie, the client IP address) from the client's DNS traffic. A strength of \system{} is that it protects the operator of the recursive resolver itself from the adversary, rather than simply the client. Conversely, existing DNS privacy mechanisms, such as DNS-over-TLS or DNS-over-HTTPS, allow for linkability between client IP and query traffic at the recursive resolver. 

Ultimately, we aim to achieve similar goals to privacy-focused DNS systems such as Quad9. However, our goals are stricter as we wish to prevent any entity in the global DNS infrastructure from associating client identity and query traffic. 

The primary goal of~\system~is to disallow the recursive resolver to have access to both the client IP address and the DNS query information. \system{} must decouple these two pieces of information by hiding the client's DNS query traffic from the recursive resolver. If the recursive resolver is unable to know the DNS queries that a client has issued, the DNS operator is unable to provide an adversary with the requested data and it will be able to remain oblivious to the requests and responses it serves.

Conversely, the \system{} resolver fundamentally has access to client queries. Therefore, we must prevent it from having access to the client IP address. 
\section{Oblivious DNS (\system{})}
\label{sec:design}
This section describes the design of \system{}. \system{} prevents an adversary
from linking a DNS query with the client IP address that issued the query. We assume
that an adversary can (1)~request data (using a subpoena or warrant) from any number of DNS operators; or (2)~access data and logs (\eg, query logs) at any DNS server.

\begin{figure}[t]
    \centering
    \includegraphics[width=.475\textwidth]{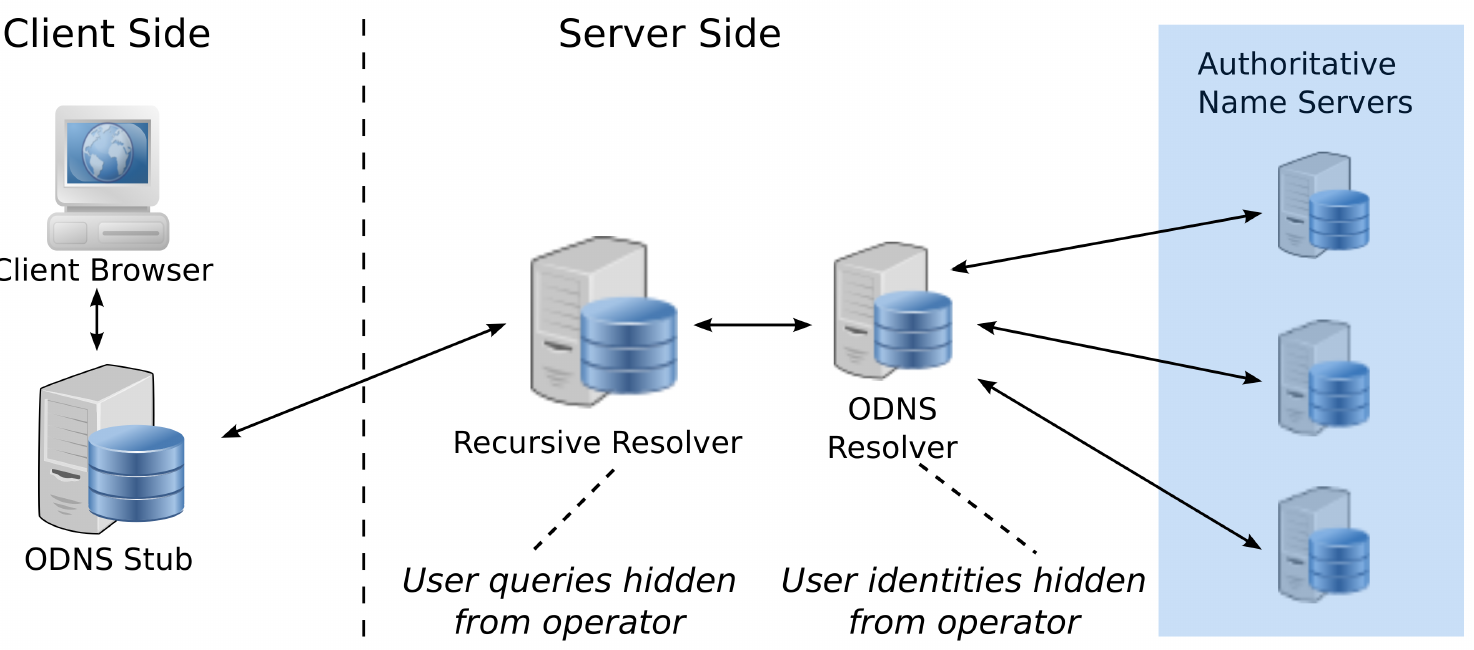}
    \caption{Overview of interacting components in \system{}.}
    \label{fig:odns_overview}
\end{figure}

\subsection{\system{} Overview}

Figure \ref{fig:odns_overview} summarizes the \system{} design. \system{} operates
similarly to conventional DNS, but alters two components: (1)~each client runs a modified stub resolver; and (2) an organization operating \system{} runs an authoritative name server (\system{} resolver) that also acts as a recursive DNS resolver for the original DNS query.

Operators of recursive resolvers see both individual client IP addresses and domains in queries. Operators of \system{} resolvers may also be able to learn information about the client if the recursive uses the EDNS0 client subnet extension. As mentioned in Section~\ref{sec:background}, EDNS0 client subnet can reveal information about the client's subnet to authoritative DNS servers higher in the DNS hierarchy (not only recursive DNS resolvers).

The recursive DNS resolver has access to the client IP address, but it never sees the
domains that it queries. \system{} requires the client to use a custom local stub
resolver, which hides the requested domain from the recursive resolver. The \system{}
stub resolver encrypts the original DNS query and the key used for encryption before it appends a plaintext \system{}-specific domain (\eg, {\tt .odns}) to the query, which causes the recursive resolver to forward the encrypted domain name on to the appropriate authoritative server (an \system{} resolver). The recursive DNS resolver receives the request from the client stub, but cannot identify the genuine domain. It queries the appropriate TLD nameserver and forwards the request to the \system{} resolver.

When an \system{} resolver receives a DNS query, it (1) decrypts the symmetric key used to encrypt the domain; and (2) decrypts the domain with the symmetric key; and (3) acts as a conventional recursive resolver for the decrypted domain.\footnote{For simplicity, we say that this authoritative server is for \url{.odns} domains, but \system{} could run on any DNS domain, and any organization that currently operates authoritative nameservers could run their own \system{} resolvers for their domain(s).} Once an answer is obtained, the \system{} resolver encrypts the response using the session key and places it in a resource record known as an OPT~\cite{RFC6891} in the response. The \system{} resolver returns the response to the original recursive DNS resolver, which in turn sends the response to the client. As explained by the use of session keys, the recursive resolver cannot learn the domains a client requests, despite being able to learn who the client is.

\subsection{Sending and Receiving ODNS Queries}
\label{sec:protocol_requests}
\begin{figure}[t!]
\centering
\includegraphics[width=.5\textwidth]{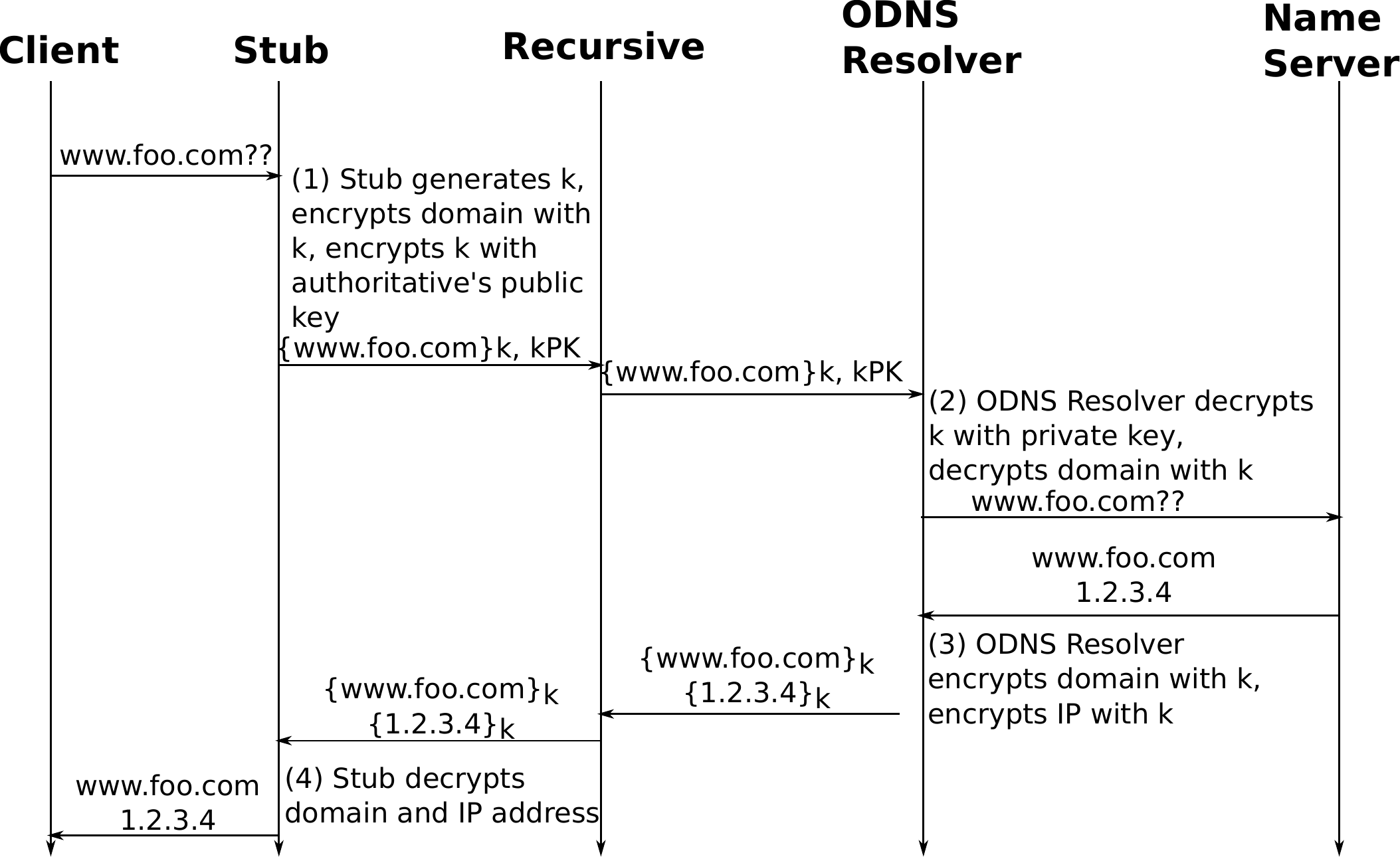}
\caption{\system{} protocol.}
\label{fig:protocol}
\end{figure}

Figure \ref{fig:protocol} shows the steps involved in answering a client's DNS request, which proceed as follows:
\begin{enumerate}
\item When a client issues a DNS request, the local stub resolver generates a symmetric session key, encrypts the domain name with the session key, encrypts the session key with the \system{} resolver's public key, and appends the \system{} domain (\eg, {\tt .odns}) to the encrypted query. ($\{$\url{www.foo.com}$\}_k$\url{.odns}.) The stub also appends the session key encrypted under the \system{} resolver's public key ($\{k\}_{PK}$)
\item The stub sends the query to the recursive resolver, which then sends it to the authoritative nameserver for the specified \system{} domain.
\item The \system{} resolver decrypts the session key, which it then uses to decrypt the genuine domain in the query.
\item The \system{} resolver forwards a recursive DNS request to the appropriate name server for the plaintext domain, which then returns the answer to the \system{} resolver.
\item The \system{} resolver returns an encrypted answer to the client's recursive resolver.
\end{enumerate}
\noindent
Other (non-\system{}) authoritative DNS servers see incoming DNS requests, but these only see the IP address of the \system{} resolver, which effectively proxies the DNS request for the original client. The client's original recursive resolver can learn the client's IP address, but cannot learn the domain names in the client's DNS queries. An additional threat is similar to the issue of the Tor edge-nodes: upstream DNS servers from the \system{} resolver can associate and aggregate the plaintext DNS requests with the IP address of the \system{} resolver and glean information from that. Mitigation is discussed in Section~\ref{sec:challenges}.

\subsection[Replication and Privacy-Preserving Key Distribution]{Replication and Privacy-Preserving\\Key Distribution}\label{sec:keys}

The \system{} resolver performs the role of both a typical authoritative server and a recursive resolver. Thus, the \system{} resolver may face higher traffic volume than a traditional authoritative nameserver. Additionally, the \system{} resolver is responsible for, and must expect, queries for pseudorandom hostnames within its zone. Thus, the \system{} resolver may be susceptible to DDoS attacks as it must attempt to decrypt all queries that it receives. We must thus design \system{} to handle the possibly high query volumes.

\textbf{Scalability and performance using anycast replicas. } In order to achieve scalability for the \system{} resolver servers, we expect to replicate them across several instances, both in geographically different locations as well as within datacenters. All replicas are assigned to both an anycast IP address as well as a unique unicast IP address. Using anycast, all servers that share the IP address are able to answer a query. The anycast address is advertised to the TLD server as the nameserver (NS) records for the \system{} domain. When a recursive resolver sends a DNS query to the \system{} resolver, the query will be routed by BGP to an server that is nearby (according to wide-area routing). Additionally, because the recursive resolver (an open resolver) often also uses anycast, both the recursive and the \system{} resolver should be the best choices based on the client's network connectivity {\it without revealing the client's location}. 
  
\textbf{Key distribution. } Use of anycast and multiple authoritative replicas introduce a key distribution challenge. Recall that the \system{} stub server uses the public key of the \system{} resolver to encrypt session keys in \system{} queries. Based on best practices, we cannot share public / private keypairs across all of the replicated \system{} resolvers. Likewise, to preserve user privacy the key distribution must be done in a way such that the \system{} resolver never learns the identity (\ie, IP address) of a client. This disqualifies out-of-band key exchange as proposed in~\cite{herrmann2014encdns}. Instead, we leverage the DNS infrastructure itself to distribute keys while maintaining privacy. We define a ``special'' query (\eg, {\tt special.odns}) that we use to select a specific \system{} resolver as well as distribute the appropriate public key. 

\begin{figure}[t!]
\centering
\includegraphics[width=.5\textwidth]{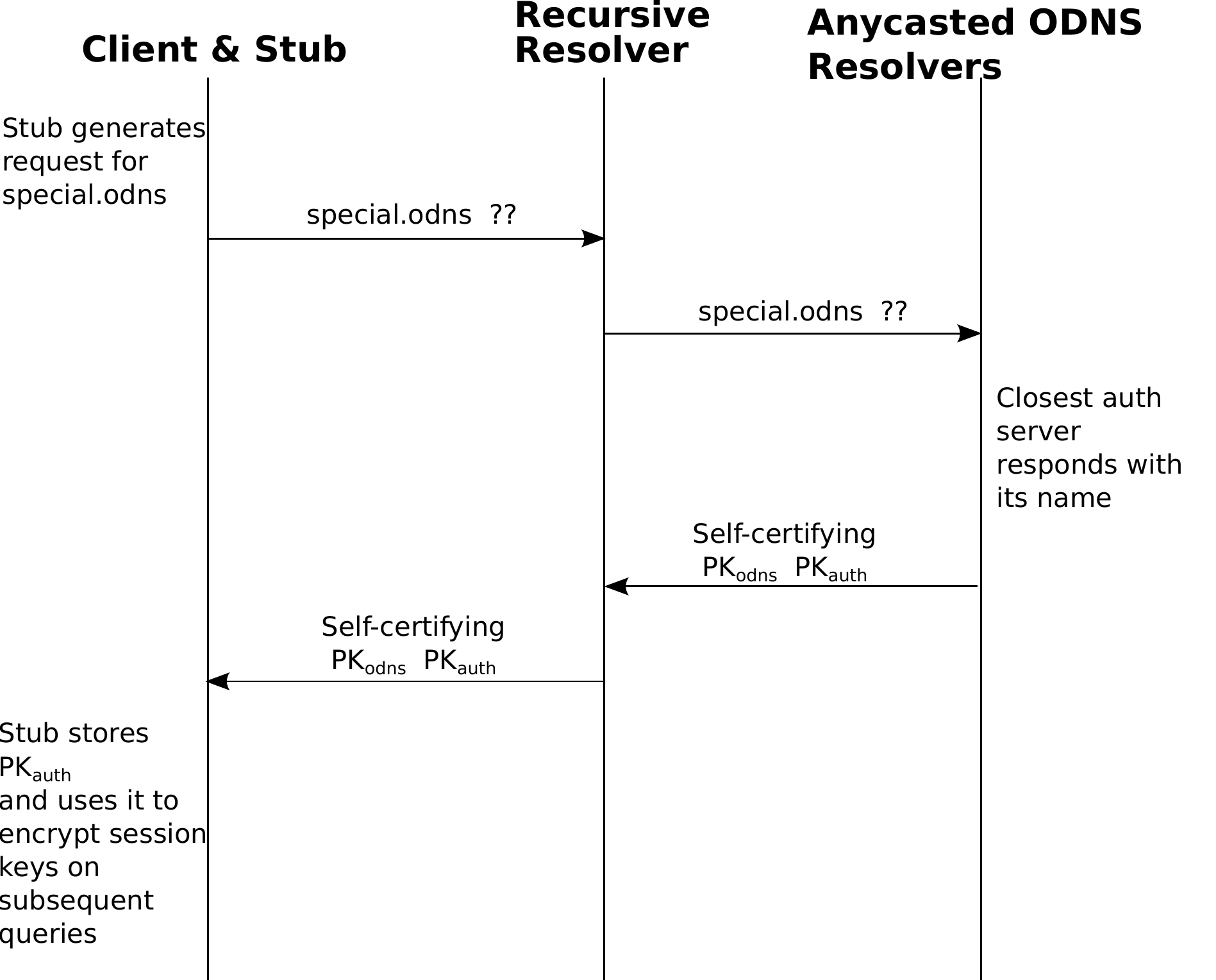}
\caption{\system{} protocol for key distribution and selecting the optimal \system{} resolver.}
\label{fig:protocol_keys}
\end{figure}
Figure \ref{fig:protocol_keys} shows this process. The client's stub resolver sends a special \system{} query to the recursive resolver, which will in turn use the anycast address to locate the nearest \system{} resolver. The \system{} resolver that receives the query responds with an OPT record that includes a self-certifying name~\cite{Mazieres,Andersen:2008}, such that the name of the server is derived from the public key itself and is associated with an instance of the \system{} resolver listening on the unique unicast IP address, and the \system{} resolver's public key. Subsequent \system{} queries at the stub append the unique name of the \system{} resolver that responded to queries, which means that the requests will all reach the same server and the client encrypts queries using the appropriate public key. Anycast servers are impacted by BGP routing updates. However, prior work~\cite{Calder:2015:APA:2815675.2815717} showed that BGP updates caused relatively low churn for clients connecting to globally-replicated anycast hosts. Based on this, clients should perform the key distribution process at regular intervals (\eg, weekly) to update the \system{} resolver that is responsible for resolving its queries.   
\subsection{Practical Challenges}
\label{sec:challenges}
To reduce barriers to deployment, \system{} must be fully-compatible with the existing global DNS infrastructure, as changes to the DNS system occur over long time scales. Thus, we design \system{} to avoid any changes at existing recursive resolvers, root nameservers, or TLD nameservers. In this section, we illuminate several practical challenges the system must overcome.

\textbf{QNAME length. }We initially attempted to transmit our encrypted domain queries in additional resource records (RR) section of a DNS message (known as an OPT). However, in practice we discovered that recursive resolvers strip additional records. Therefore, we use the QNAME field to transmit both the encrypted domain query as well as the asymmetrically encrypted session key as the QNAME field is preserved while passing through recursive resolvers. A DNS QNAME field consists of 4 sets of 63 bytes, which limits both the key size and encryption scheme used.

\begin{figure}[t!]
\centering
\includegraphics[width=\columnwidth]{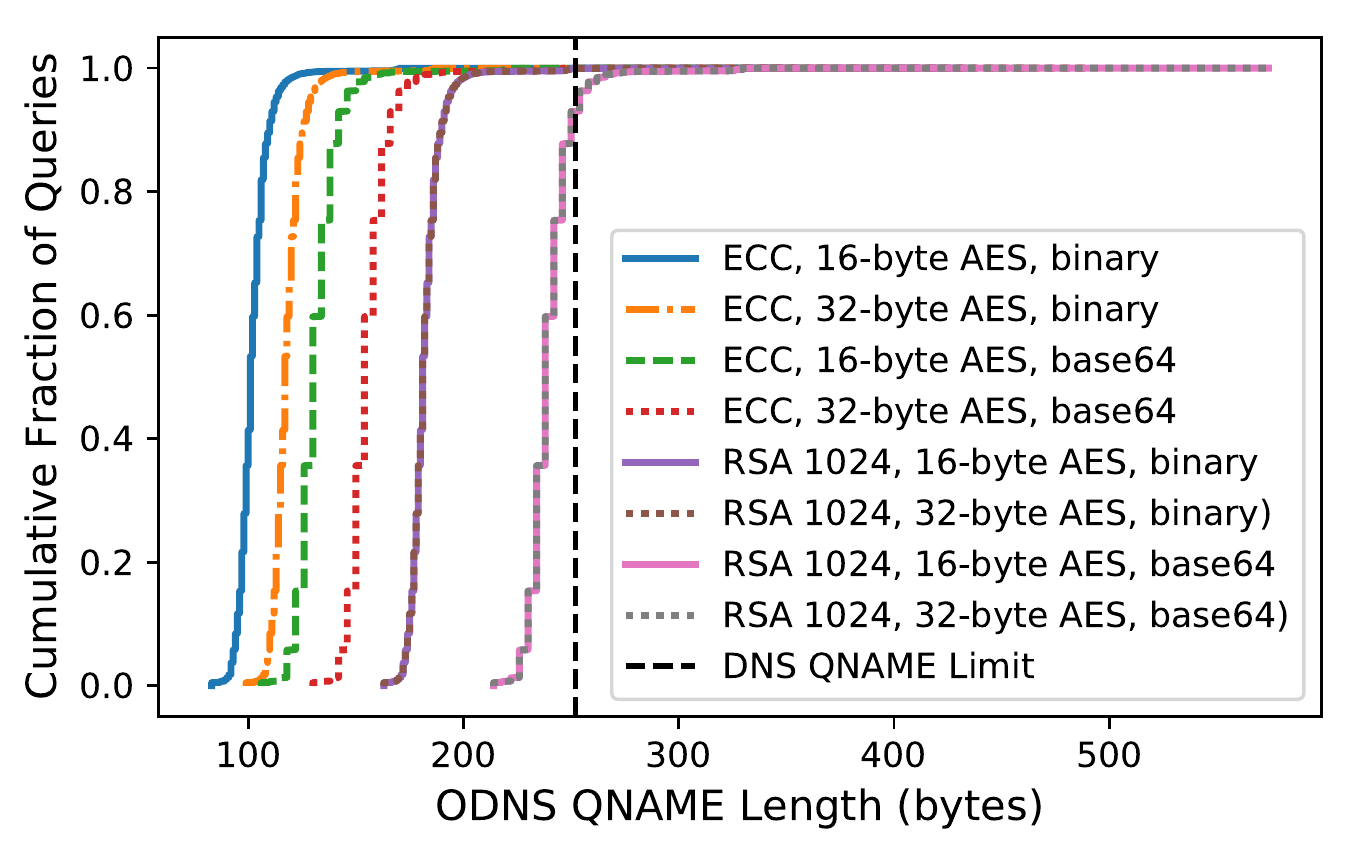}
\caption{Encrypted QNAME lengths using different cryptographic primitives.}
\label{fig:qnames}
\end{figure}

We investigate combinations of the required cryptographic primitives (symmetric AES encryption for domains, asymmetric encrypytion for symmetric keys) and the resulting QNAME lengths to assess the suitability of the QNAME field with regard to \system{}'s encryption scheme. We calculate total QNAME lengths for all query records in an example query list used for testing DNS server performance\footnote{\url{http://wwwns.akamai.com/queryfile-example-current.gz}}. The list contains 6.8M queries and we calculate the lengths using both binary and base64 encoding. Figure~\ref{fig:qnames} shows the results. We see that both 1024-bit RSA and ECC using Curve25519 tend to result in query lengths that are well below the 252-byte limit. However with the use of base64 and RSA-based asymmetric encryption the likelihood of generating an invalid QNAME is increased. For this reason, \system{} uses 16-byte AES session keys and encrypts the session keys using the Elliptic Curve Integrated Encryption Scheme (ECIES)~\cite{ecies}. We anticipate that there may be cryptographic primitives that will result in lower overhead; however, AES and ECIES libraries are readily available for our development platform. In our prototype deployment, which uses base64 encoding, we discovered that 0.6\% of the queries in the list had lengths that exceeded the QNAME length limit. 

To allow the \system{} prototype to operate in conjunction with the deployed
DNS infrastructure, we configured the prototype \system{} stub resolver to fall back to
conventional DNS lookups whenever plaintext QNAME fields that result in
ciphertext QNAMEs that are larger than the maximimum field size. To preserve
privacy, queries that result in invalid QNAME lengths will receive FORMERR
errors in the deployed version of \system{}.

\textbf{EDNS0 client subnet. }As discussed in Section~\ref{sec:background}, the EDNS0 client subnet option allows upstream authoritative nameservers to learn the IP subnet of clients issuing queries. To achieve the intended privacy benefits of \system{}, the client subnet option should not be used. We intend to require in the \system{} specification in development aimed at the IETF that \system{} stub resolvers use EDNS0 Client Subnet with the value of 0, which requires the recursive to not attach ECS for them. Assuming this were to fail (as even specification requirements are not always adhered to), the risk is that an adversary can spy the IP addresses in plaintext ECS sent from the recursive to the \system{} resolver. This risk could be mitigated if only recursives that will encrypt the traffic to authoritatives, using DNS-over-TLS or one of the other methods, were selected by the stub. Likewise, \system{} resolvers can readily observe whether ECS has been attached to a query or if encrypted transport between the recursive and the authoritative is in-use and embed a client notification in the subsequent response. We leave this functionality for future work.

To better understand real-world usage of the EDNS0 client subnet option, we
performed a short experiment to discover publicly available recursive resolvers
that have the option enabled. We operate authoritative nameservers for the
domain {\tt \url{obliviousdns.com}} and collect pcaps of queries that reach the servers. Queries are initiated by using \texttt{dig} against a
list of 16 popular publicly available recursive resolvers. Overall, we find
that only four of the sixteen resolvers implement EDNS0 client subnet, as
shown in Table~\ref{tab:realdns}. Several popular recursive resolvers do not
currently implement EDNS0 client subnet (such as Cloudflare's {\tt 1.1.1.1} or Quad9's
{\tt 9.9.9.9}).

\textbf{0x20-bit encoding. }0x20-bit encoding is a scheme intended to
thwart DNS poisoning attacks by altering the QNAME field in queries to a mix
of uppercase and lowercase (\eg, {\tt \url{www.foo.com}} to
{\tt \url{wWW.FoO.CoM}})~\cite{0x20}. This technique increases the difficulty of
cache poisoning as the attacker must guess the mixed-case QNAME. The \system{}
prototype encodes the encrypted domain and session keys using base64, which
results in a character set including both upper and lowercase letters.
Therefore, for our purposes, the recursive resolver between the \system{} stub
and the \system{} resolver must not use 0x20-bit encoding, as it would
render decryption impossible. Table~\ref{tab:realdns} shows the public
recursive resolvers that currently implement 0x20. Note that 0x20-bit encoding
does not pose a challenge for binary-encoded queries, which are
commonly supported in real-world infrastructure~\cite{herrmann2014encdns}.
\begin{table}[t]
\centering
\resizebox{.9\columnwidth}{!}{
\begin{tabular}{|l|c|c|}
\hline
\textbf{Open Recursive Resolver (IP)}&\textbf{EDNS0 Client Subnet}&\textbf{0x20}\\ \hline
Google (8.8.8.8)&\cmark& \\ \hline
Dyn (216.146.35.35)&\cmark& \\ \hline
Fourth Estate (45.77.165.194)&\cmark& \\ \hline
GreenTeamDNS (81.218.119.11)&\cmark& \\ \hline
Cloudflare (1.1.1.1)& &\cmark\\ \hline
Verisign (64.6.64.6)& &\cmark\\ \hline
Quad9 (9.9.9.9)& & \\ \hline
Level3 (209.244.0.3)& & \\ \hline
OpenDNS Home (208.67.222.222)& & \\ \hline
Norton ConnectSafe (199.85.126.10)& & \\ \hline
Comodo Secure DNS (8.26.56.26)& & \\ \hline
DNS.WATCH (84.200.69.80)& & \\ \hline
SafeDNS (195.46.39.39)& & \\ \hline
FreeDNS (37.235.1.174)& & \\ \hline
Hurricane Electric (74.82.42.42)& & \\ \hline
Ultra (156.154.71.1)& & \\ \hline
\end{tabular}
}
\caption{Open recursive resolvers with EDNS0 client subnet and 0x20-bit encoding enabled.}
\label{tab:realdns}
\end{table}

\textbf{Caching. }\system{} caches queries at the \system{} resolver. As the \system{} resolver is essentially an extra recursive resolver in the system, the \system{} resolver simply acts as the shared cache, in lieu of the shared cache that would ordinarily exist at the ISP recursive resolver. This pushes caching further into the DNS hierarchy, decreasing the time to conduct a look up for the domain from the perspective of the authoritative server. In addition to caching at the \system{} resolver, \system{} can also cache DNS responses at the stub resolver; while this may provide some caching benefits on a per-client basis, it does not provide cross-client caching benefits. It is important to note that caching at the recursive resolver would not provide any performance enhancement for the clients in this system. Each DNS query is encrypted with a new session key $k$, thus two DNS queries for the same domain do not appear the same to these resolvers (\{{\tt \url{www.foo.com}}\}$_{k1}$ $\neq$ \{{\tt \url{www.foo.com}}\}$_{k2}$); therefore, if the recursive resolver cached the response for \{{\tt \url{www.foo.com}}\}$_{k1}$, it would never see a cache hit for that entry because subsequent lookups for {\tt \url{www.foo.com}} appear as a different URL. 

We use two techniques to avoid causing the recursive resolver to cache \system{} responses (thus potentially ejecting conventional cached entries). First, our \system{} resolver sets the TTL value of the response to zero. The TTL value is meant to indicate the number of seconds a DNS response should be considered valid. By setting the TTL to zero, we essentially indicate that all \system{} responses should not be cached. Additionally, we send the response in an OPT record, which should not be cached by recursive resolvers. We further explore the implications on caching and traffic at the recursive resolver in Section~\ref{sec:caching}. 

\textbf{\system{} ingress / egress timing vulnerability. }An adversary with sufficient visibility might be able to view both the encrypted \system{} queries and the subsequent plaintext queries originating from the \system{} resolver. Such a case could also arise if both the stub and the \system{} resolver use the same recursive resolver for lookups. \system{} can avoid these issues in three straightforward ways: (1)~the \system{} resolver can inspect the source of incoming queries and maintains rules to avoid using the same service for the plaintext query; (2)~\system{} resolvers can simply repeat the same actions as the stub by encrypting the encrypted query again (\ie, behavior akin to Tor); and (3)~send the queries from the \system{} server using query name minimization~\cite{RFC7816} so that root and popular TLD operators do not aggregate the history of the queries from each \system{} server, because this is an additional correlation risk.

\textbf{.odns top-level domain. }In our examples we discuss the use of a top-level domain specifically for the use of \system{}. In reality, TLDs are challenging and costly to procure, requiring extensive time and effort. Due to these barriers, we have implemented and evaluated our prototype while using the registered domain {\tt \url{obliviousdns.com}}. This long DNS domain currently consumes more of the already-limited QNAME field with our appended domain name. Widespread adoption of \system{} could lead us to acquire a new TLD in the future or select a shorter \system{}-specific domain. As previously stated, organizations that own existing domains can deploy \system{} using their own authoritative namespace.
\subsection{Implementation}
\label{sec:implementation}
For experimentation and prototyping, we implement a prototype of the \system{} stub and the \system{} resolver in Go. We built this functionality as extensions to an existing Go DNS library~\cite{miekg}. This implementation choice allowed us to rapidly prototype our design and to quickly test various features. We anticipate a future implementation would be built open popular DNS resolvers such as Knot or Unbound.

We use a hybrid encryption scheme and leverage well-known cryptographic primitives in ODNS to balance privacy with storage requirements. For each \system{} request, the stub generates a 16-byte AES symmetric key. This key encrypts the plaintext request from the user. The AES session key is then asymmetrically encrypted using the \system{} resolver's public key with ECIES~\cite{ecies} using Curve25519~\cite{curve25519}. Elliptic-curve cryptography allows \system{} to reap the performance benefits of elliptic curves and take advantage of the smaller key sizes needed to provide equivalent security as other public-key implementations.
\section{Evaluation}
\label{sec:eval}
In this section, we study the performance of \system{} in terms of both (1)~the overhead on individual lookups; (2)~the impact of using \system{} when loading many popular websites; and (3)~the implications of using \system{} on existing infrastructure in terms of caching. 

\subsection{Microbenchmarks}
\label{sec:micro}
In this section we explore the overhead introduced by \system{} on individual lookups.

\begin{figure}[t!]
\centering
\includegraphics[width=.475\textwidth]{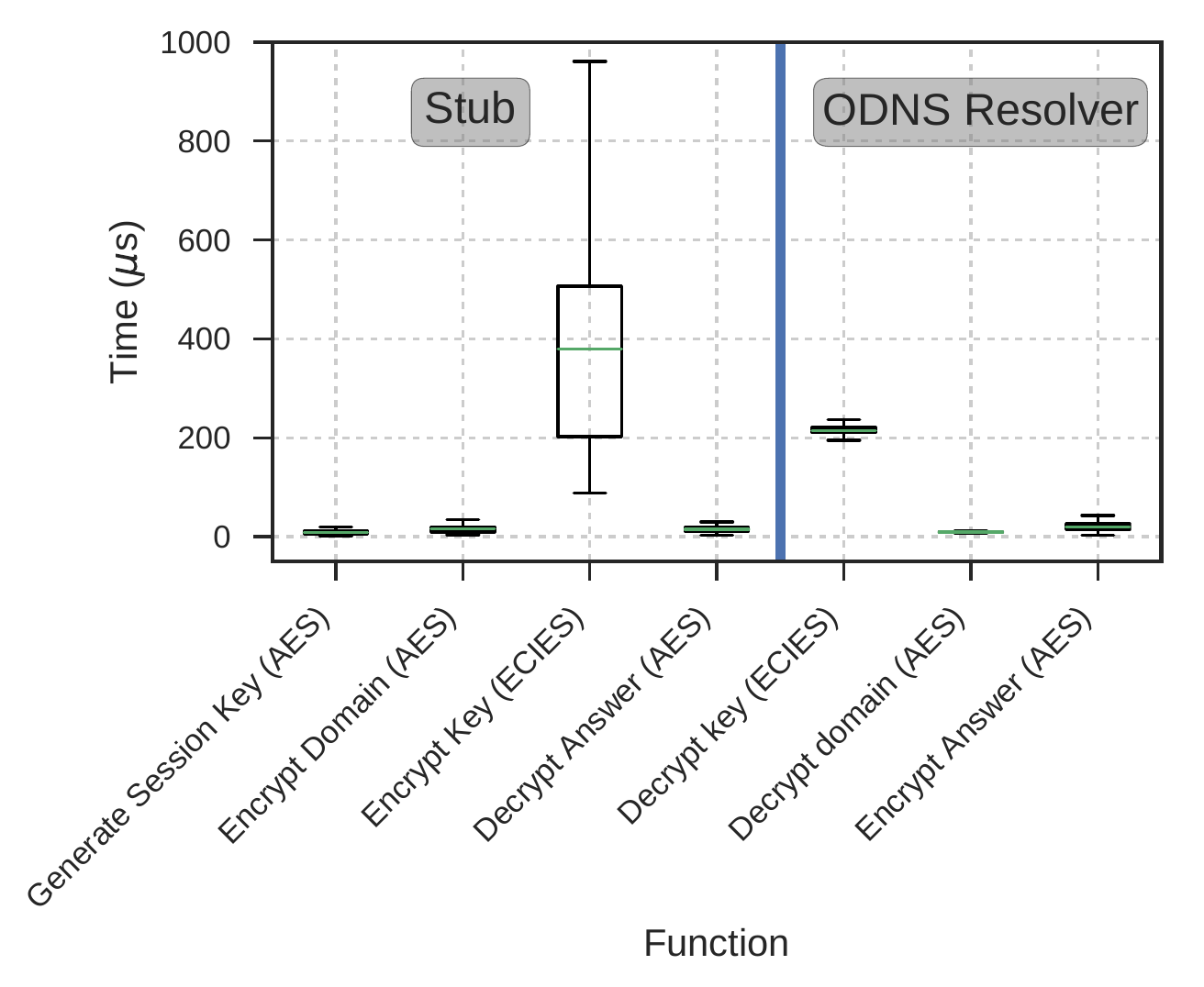}
\caption{Overhead of different cryptographic operations performed in the \system{} protocol.}
\label{fig:overhead_parts}
\end{figure}
\textbf{Cryptographic overhead. }We first investigate the overhead introduced by each of the cryptographic operations that occur in an \system{} lookup. We benchmark the individual operations performed in the \system{} protocol to attribute overhead to different parts of the protocol. We run a DNS query list composed of the Alexa Top 10,000 domains and record the duration of each operation. The results of these microbenchmarks are shown in Figure~\ref{fig:overhead_parts}. We see that the symmetric operations using AES result in relatively little overhead at both the stub and the \system{} resolver. The asymmetric operations using ECIES are more costly and contribute to the majority of the total overhead of \system{}. However, even the most costly operation, asymmetrically encrypting the AES key using ECIES, takes less than 1 ms due to the relatively high performance of elliptic-curve cryptography. Overall, the sum of all cryptographic operations, including at both the stub and the \system{} resolver, looks to take roughly 1 ms. Additionally, we anticipate further improvements are attainable using optimized cryptographic libraries, whereas our prototype uses standard libraries.

\begin{figure}[t!]
\centering
\includegraphics[width=.425\textwidth]{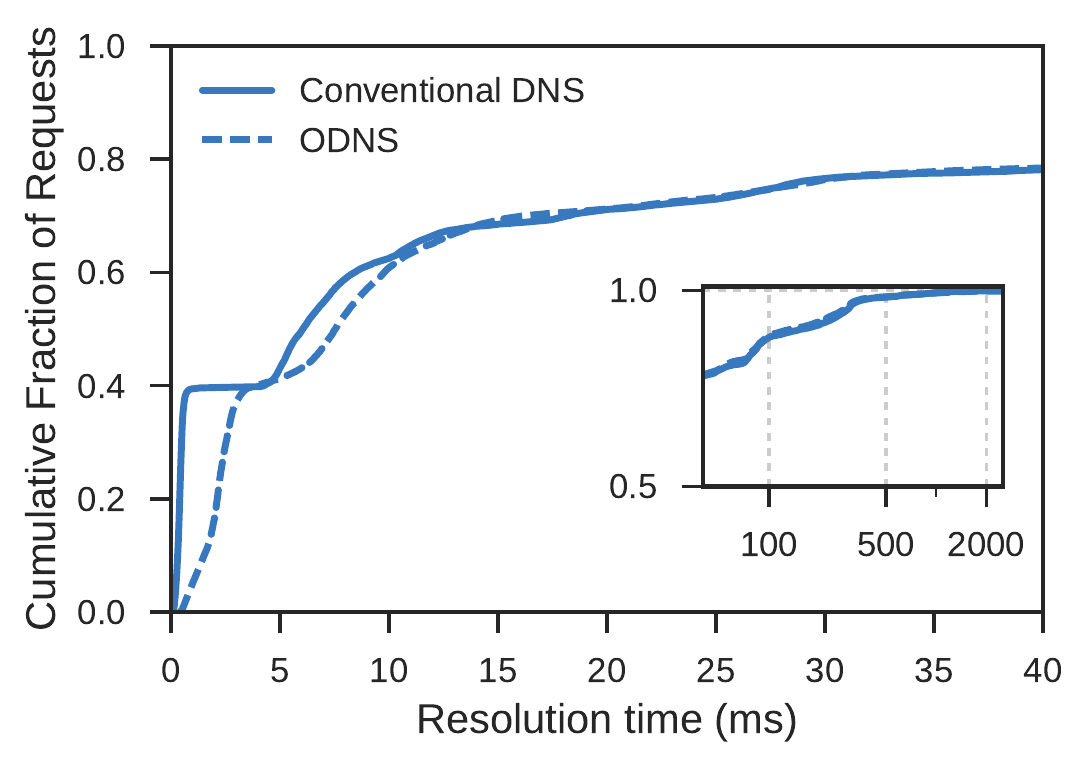}
\caption{Additional latency running \system{} compared with conventional DNS.}
\label{fig:overhead_local}
\end{figure}

\textbf{\system{} protocol overhead. }We compare the total latency introduced by \system{} for each query to that of conventional DNS by using the {\tt dig} command to issue DNS queries. To single out \system{} protocol overhead that isn't impacted by network latency between servers, we install both the \system{} stub and resolver onto a single machine. We query the Alexa Top 10,000 domains with \system{}'s functionality both enabled and disabled to compare lookup latency distributions. The results of this are shown in Figure~\ref{fig:overhead_local}; we can see that the use of \system{} results in longer query resolution times, but generally follows the same trend as conventional DNS. Overall, the median resolution time for conventional DNS is 6.03 ms, while the median time for \system{} resolutions is 7.53 ms. The overhead difference corroborates the cryptographic overhead witnessed above. The additional latency can be attributed to the stub and \system{} resolver process functions.

\begin{figure}[t!]
\centering
\includegraphics[width=.45\textwidth]{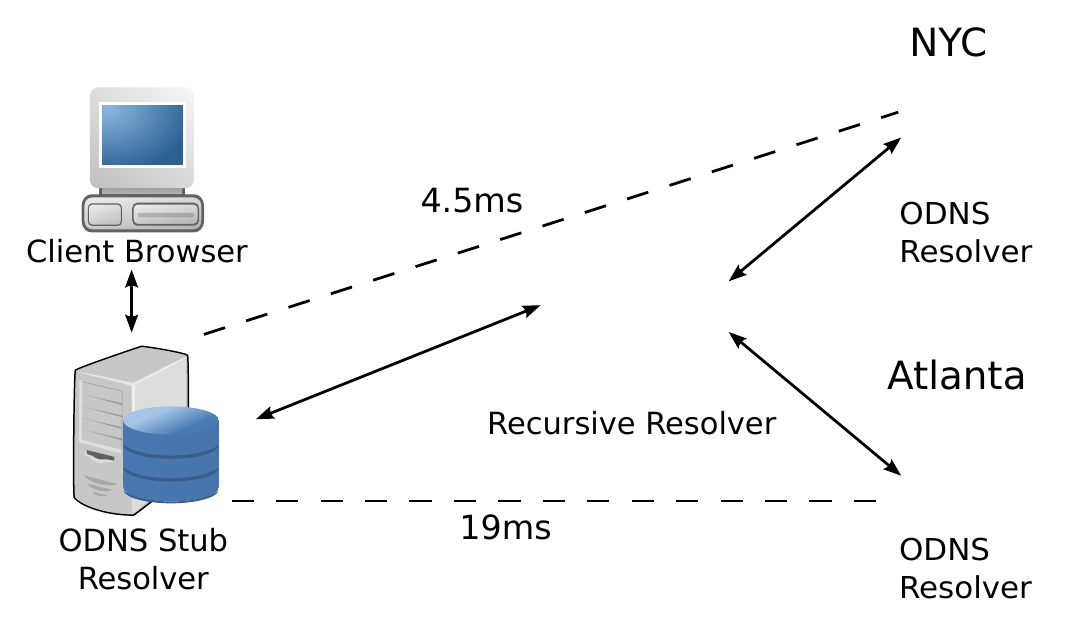}
\caption{Deployment Setup.}
\label{fig:prototype}
\end{figure}

\textbf{Latency in the wide-area. }To test \system{} performance in the wide-area, we deploy our prototype on real-world infrastructure. Figure~\ref{fig:prototype} shows our deployed evaluation setup. We test \system{} using a client in the New York City metropolitan area, running the \system{} stub resolver running on the same local machine, communicating with its default recursive resolver. 

We evaluate \system{}'s query overhead to that of conventional DNS by issuing DNS queries for the Alexa Top 10,000 domains using both conventional recursive resolvers ({\tt 1.1.1.1}, {\tt 9.9.9.9}, and {\tt 8.8.8.8}) and \system{} with two \system{} resolvers---functionally equivalent---in two different locations: Georgia and New York City. Recall that the client and stub resolver are also located in the New York City area. To evaluate the effect of the \system{} resolver's geographic proximity to the client, we measure the overhead using each of the \system{} resolvers. 

\begin{figure}[t!]
\centering
\includegraphics[width=.475\textwidth]{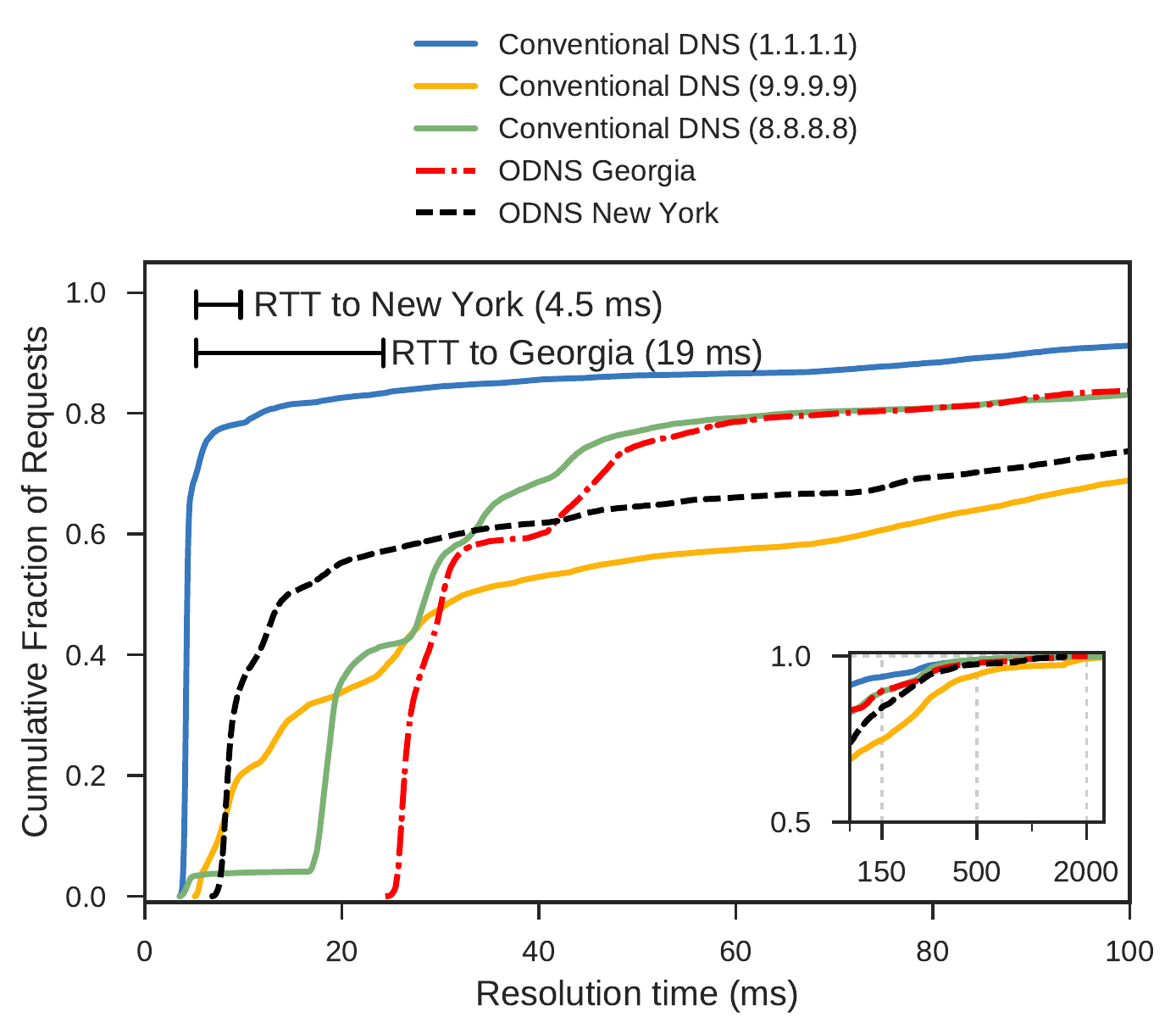}
\caption{ODNS performance with servers located in Georgia and New York.}
\label{fig:realoverhead}
\end{figure}

\begin{figure*}[t!]
\centering
\includegraphics[width=.75\textwidth,trim={10 14 12 12},clip]{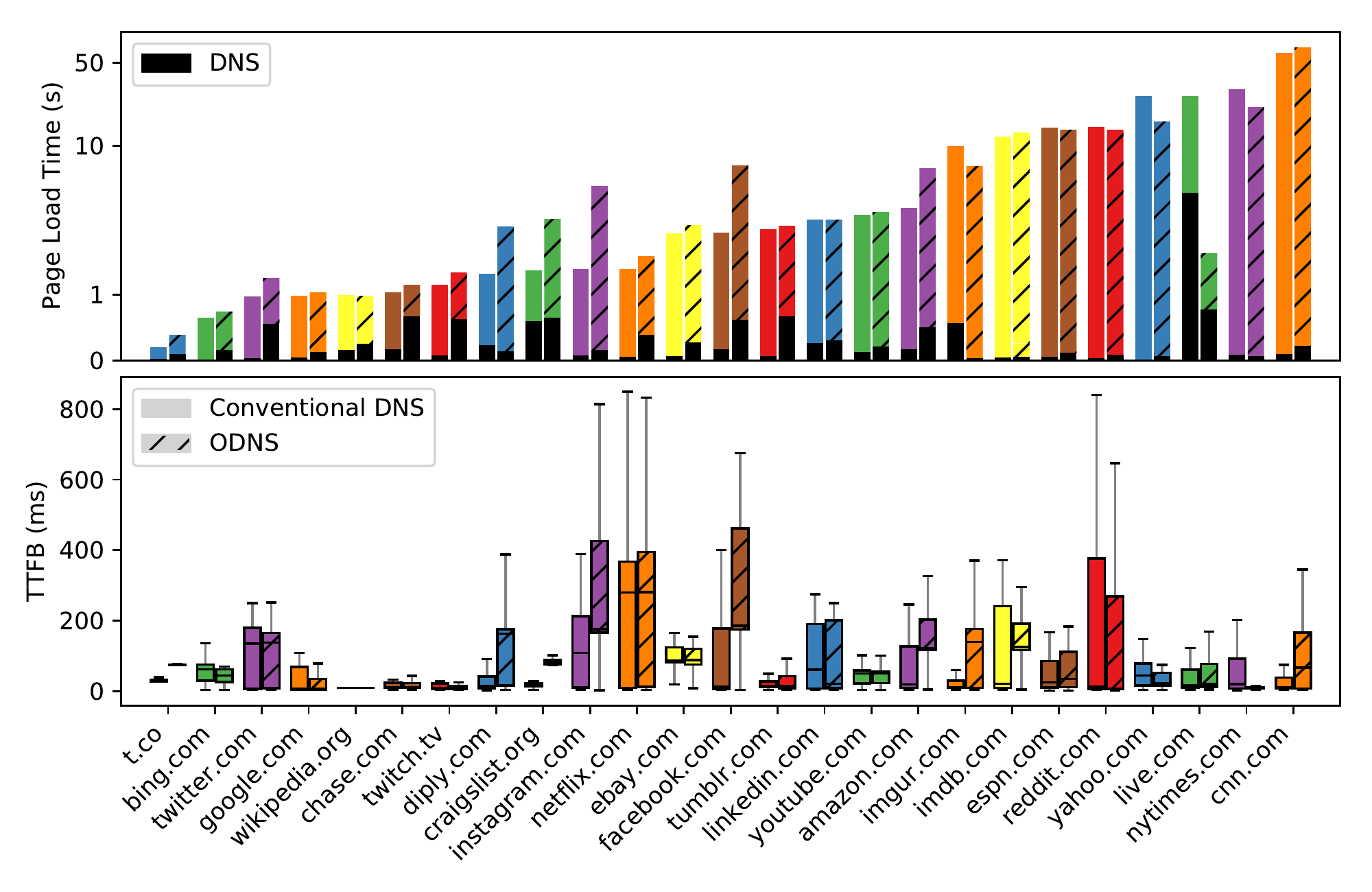}
\caption{Page load times and time-to-first-byte for various web pages using \system{} and conventional DNS.  The left bars in the figure represent conventional DNS and the right bars represent the time it takes using \system{}.}
\label{fig:page_load_time}
\end{figure*}

In this experiment, we warm a cache at the \system{} resolver using a sample domain query trace provided for use with \texttt{resperf}, a popular DNS performance testing tool. Figure~\ref{fig:realoverhead} shows the results; we can see that \system{} queries compare favorably to the popular open resolvers, and in the case of 8.8.8.8 and 9.9.9.9, \system{} tends to result in \textit{lower} latencies when using the geographically close \system{} resolver in New York. It is also evident from this experiment that the network round trip time, roughly 4.5 ms for New York and 19 ms for the Georgia server, between the stub and the \system{} resolver is added for every query. 

Given these results, \system{} does not appear to introduce significant overhead when resolving domains provided that the RTT between the client and the \system{} resolver remains reasonable. As described in Section~\ref{sec:keys}, \system{} resolvers are replicated and use anycast IP addresses in order to provide scalability and achieve desirable latencies. To get a sense for the size of an anycast deployment that would provide adequate performance, Schmidt et al.~\cite{Schmidt17a} investigated the performance of real-world anycast systems and determined that a system deployed in only 12 locations around the world can result in relatively low latency for a global client base. Though the target RTTs for DNS would likely call for a larger anycast deployment, we anticipate that a widely-deployed \system{} anycast network could achieve acceptable lookup latencies for any user. We leave investigation into the necessary anycast deployment for future research.

\subsection{Macrobenchmarks}
In this section, we explore aggregate \system{} overhead while using web applications, which can often trigger tens of DNS queries for a single website.

\textbf{Web page load time. }We measure how \system{} would affect a typical Internet user's browsing experience by evaluating the overhead of a full page load; a full page load consists of not only conducting a DNS lookup for the page, but also fetching the page, and conducting any subsequent DNS lookups for embedded objects and resources in the page. We fetch a selection of popular web pages from the Alexa top sites list in the United States\footnote{\url{https://www.alexa.com/topsites/countries/US}} using an \system{} resolver in New York as well as conventional DNS for comparison. We browse to each site 30 times using the chrome webdriver and record HAR files for each browsing session. We then calculate the mean values for page load time and DNS time for each site. The top graph of Figure~\ref{fig:page_load_time} shows the results. The left bars (unhatched) in the figure represent total page load time using conventional DNS, and the right bars (hatched) represent the time it takes using \system{}. The portion of each bar that is shaded in black illustrates the portion of page load time that is attributable to DNS lookups. We see that there is generally not a significant difference in page load time between \system{} and conventional DNS because DNS lookups contribute relatively little time to the entire page load process. In some cases, \system{} load times are actually shorter than loading a page with conventional DNS\footnote{We investigated the performance for live.com to understand how \system{} could drastically outperform conventional DNS. It appears that pages themsolves were rather different when using \system{} versus conventional DNS. We found that \system{} usage resulted in loading 7 fewer javascript objects from a different \texttt{optimizely.com} CDN. It appears that the \system{} runs were directed to a cached bundle of javascript, whereas conventional DNS runs were directed to individual javascript objects from a different optimizely CDN location.}. Given these overall results, we believe that the end-user web experience will not be greatly impacted using \system{}, presuming that there exists a nearby \system{} resolver.

\textbf{Web page time-to-first-byte. }CDNs distribute content to many locations
to achieve low latency for users as they connect to nearby copies of requested
content. A concern when using \system{} is that, for a given lookup, the location of the \system{} resolver will appear to be the location of a user, thus localization techniques may result in users connecting to CDN servers that are suboptimal. We investigate the time-to-first-byte (TTFB) values for every object on each site from our traces. This value illustrates the difference in terms of distance to content when using \system{} versus conventional DNS. The bottom graph in Figure~\ref{fig:page_load_time} shows the TTFB distributions for each site, with conventional DNS on the left and \system{} on the right. These results help explain the page load times we witness for several sites. For example, \texttt{reddit.com} and \texttt{nytimes.com} both achieve lower TTFBs using \system{}, indicating that they were directed to content servers that are closer than those resolved with conventional DNS. Accordingly, the \system{} page load times for those sites were faster when using \system{}. Likewise, when \system{} results in higher TTFBs such as with \texttt{instagram.com}, we see longer page load times. This insight motivates widespread deployment of \system{} resolvers and the use of anycast. Such techniques should allow for both low latencies to the resolvers as well as maintaining content localization benefits that CDNs can provide.

\subsection{Impact on Existing Recursive \\DNS Infrastructure}\label{sec:caching}
A critical performance aspect of conventional DNS is caching---specifically caching DNS responses at the recursive resolver. As discussed, \system{} usage results in the inability to cache usable responses at the recursive resolver as each query using a different symmetric key will be unique. Caching plays a critical role in the overall performance of DNS; therefore, we simulate the potential DNS traffic implications at a recursive resolver with varying levels of \system{} usage. 

\textbf{Cache misses. }For the first experiment, we seek to understand the potential increase in cache misses at the recursive resolver, which will result in increased traffic to nameservers located above the recursive resolver in the DNS hierarchy. Intuitively, if \system{} results in a significant DNS traffic increase at the recursive resolver, widespread usage may be discouraged by recursive resolver operators. We simulate a caching recursive resolver as well as a user population using real-world traces collected in October 2017 from a fiber-to-the-home network in Cleveland, Ohio~\cite{CCZ}. The trace consists of roughly 8 million type A DNS queries from anonymized clients over the course of one month. All clients are assigned to a single simulated recursive resolver. For each test run we randomly assign a fixed percent of the user population to be \system{} users. The ISP, as well as all \system{} stubs, include caches that have no size limits. Non-\system{} stubs do not have a cache.
\begin{figure}[t!]
\centering
\includegraphics[width=.875\columnwidth]{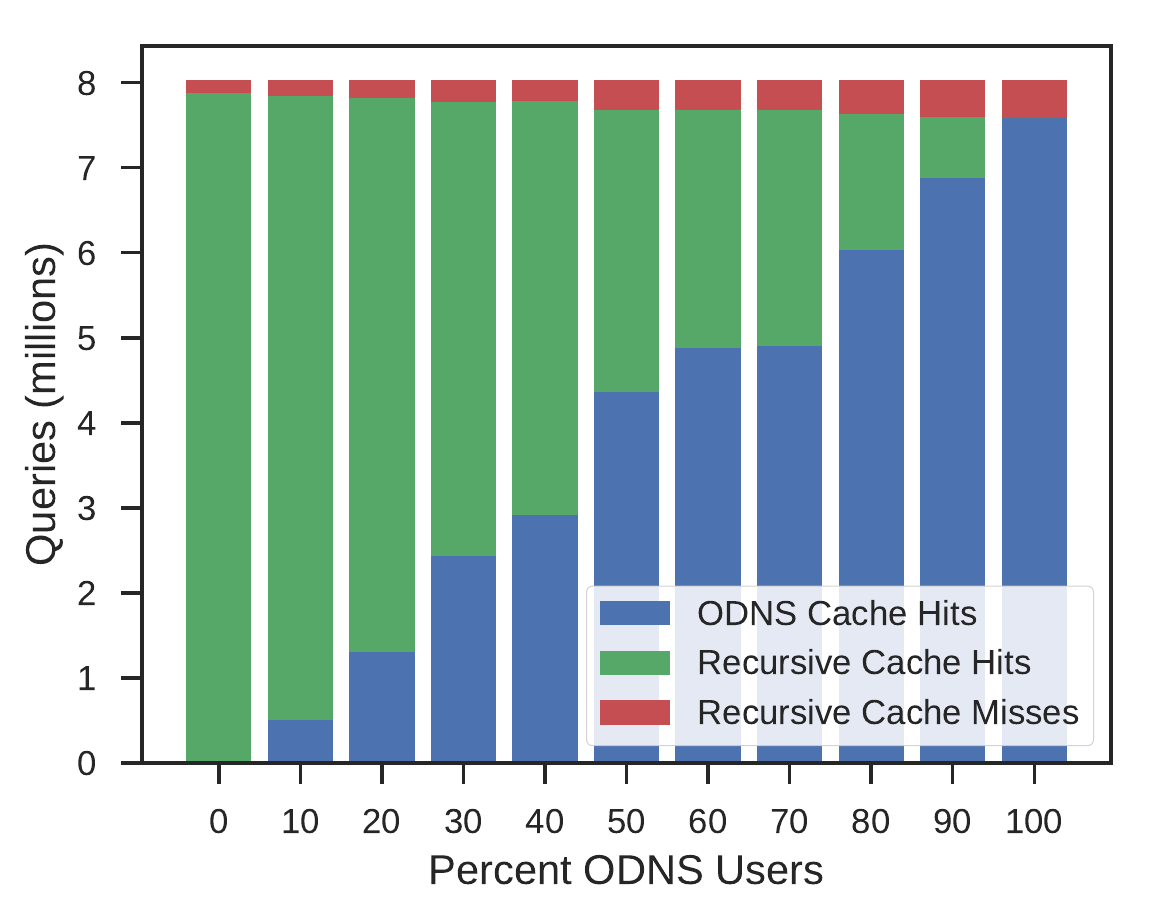}
\caption{Cache hit rates at the \system{} resolver as a function of the fraction of Internet users who use \system{}.}
\label{fig:ISPusage}
\end{figure}

The results of the simulation are shown in Figure~\ref{fig:ISPusage}. Note that we do not show ODNS cache misses as they result in recursive hits or misses. We see that \system{} usage causes a relatively modest increase of cache misses at the recursive. However, we see that, as the percent of \system{} users increases, the overall DNS traffic load on the recursive resolver decreases. For instance, with a network consisting of 20\% \system{} clients, the overall number of queries that reach the recursive resolver is reduced by 16.3\%. This is due to the presence of a cache at the stub resolvers. These results are consistent with Jung et al.'s observation~\cite{jung} that caching can provide significant benefits for even a single user. Overall, our simulation shows that \system{} \textit{reduces} the traffic burden on the recursive infrastructure\footnote{Note that some stub clients (\eg, Microsoft Windows) implement caching today, while many do not. We anticipate that \system{} stub caches would not impact recursive traffic positively or negatively for such clients, while it would reduce traffic for clients that currently do not implement a cache.}

\textbf{Unwanted \system{} cache entries. }We next explore the potential burden on recursive resolver caches caused by \system{} entries. As each \system{} query is unique, a cache entry based on the encrypted query should never be hit. To combat caching of our responses, we set their TTL value to zero at the \system{} resolver. However, recursive resolvers commonly ignore TTL values and cache responses for a longer amount of time~\cite{Pang:2004:RDN:1028788.1028792}. The result of this behavior is that \system{} responses could potentially cause valid entries to be ejected from a limited-size cache. 

\begin{figure}[t!]
\centering
\includegraphics[width=.9\columnwidth]{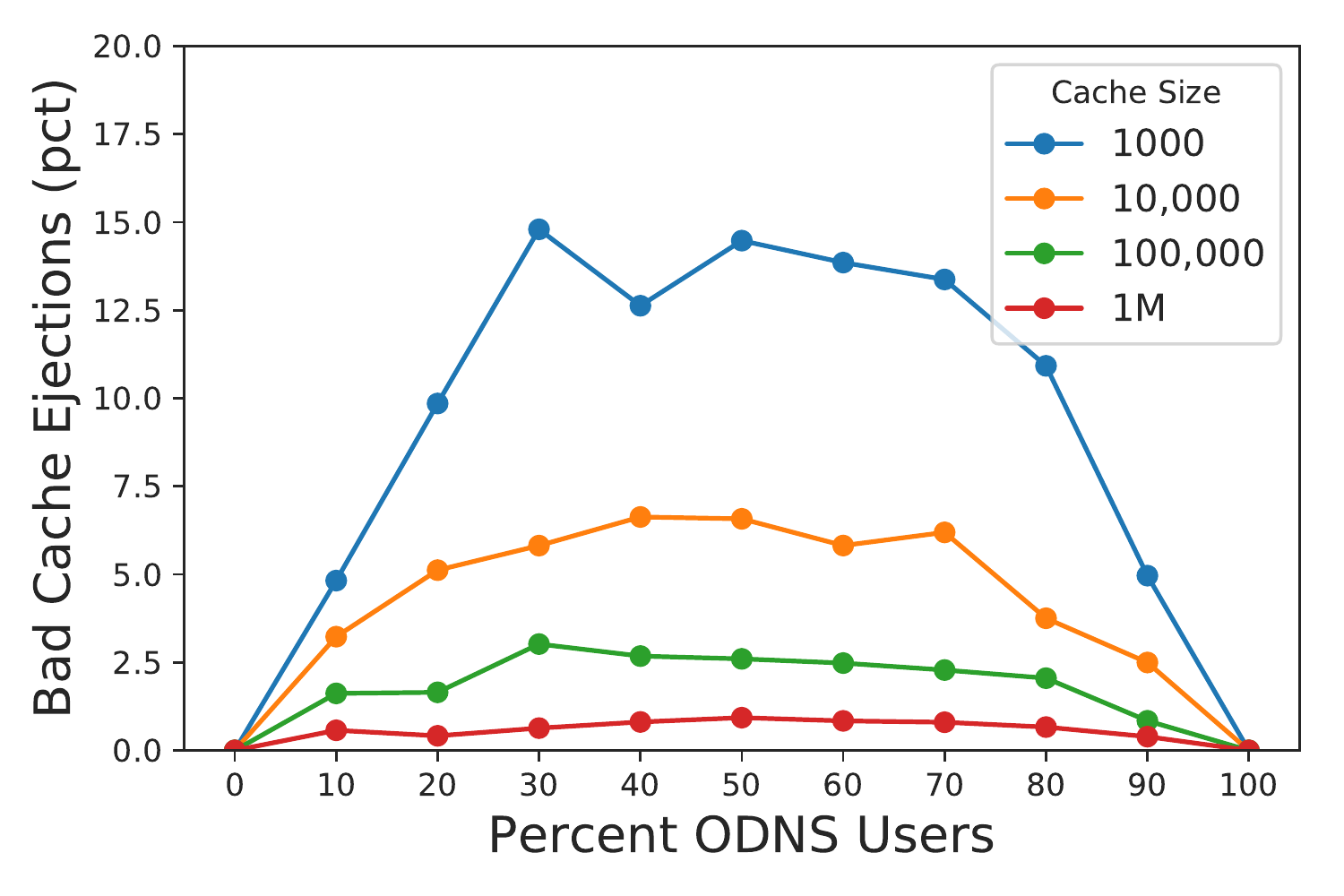}
\caption{Percent of non-ODNS records ejected from DNS server cache by ODNS records as a function of the fraction of Internet users who use ODNS.}
\label{fig:cachesize}
\end{figure}

In this simulation we use the same traces as the previous section and again vary the percent of \system{} users. We also track ``bad'' ejections, which are characterized by when an \system{} entry causes a non-\system{} entry to be ejected from a size-limited, least-recently used (LRU) cache. Figure~\ref{fig:cachesize} plots the percent of bad ejections for various cache sizes using the simulation. Overall, we observe that the percent of bad ejections tends to remain rather low; for instance, caches with 100,000 entries or more never experience more than roughly 2.6\% unwanted ejections. The percent of bad ejections decreases as usage of \system{} increases because, at higher usage levels, \system{} entries are causing other \system{} entries to be ejected. We also note that, assuming the chosen cache replacement policy is either LRU or least-frequently used, the bad ejections likely represent ejections of entries in the tail of the valid distribution and not of the most popular or common entries. These results lead to the conclusion that \system{} should not greatly impact the performance of recursive resolver caches.

\subsection{Operating the \system{} Resolver as a Recursive or a Forwarder}
\begin{figure}[t!]
\centering
\includegraphics[width=.9\columnwidth]{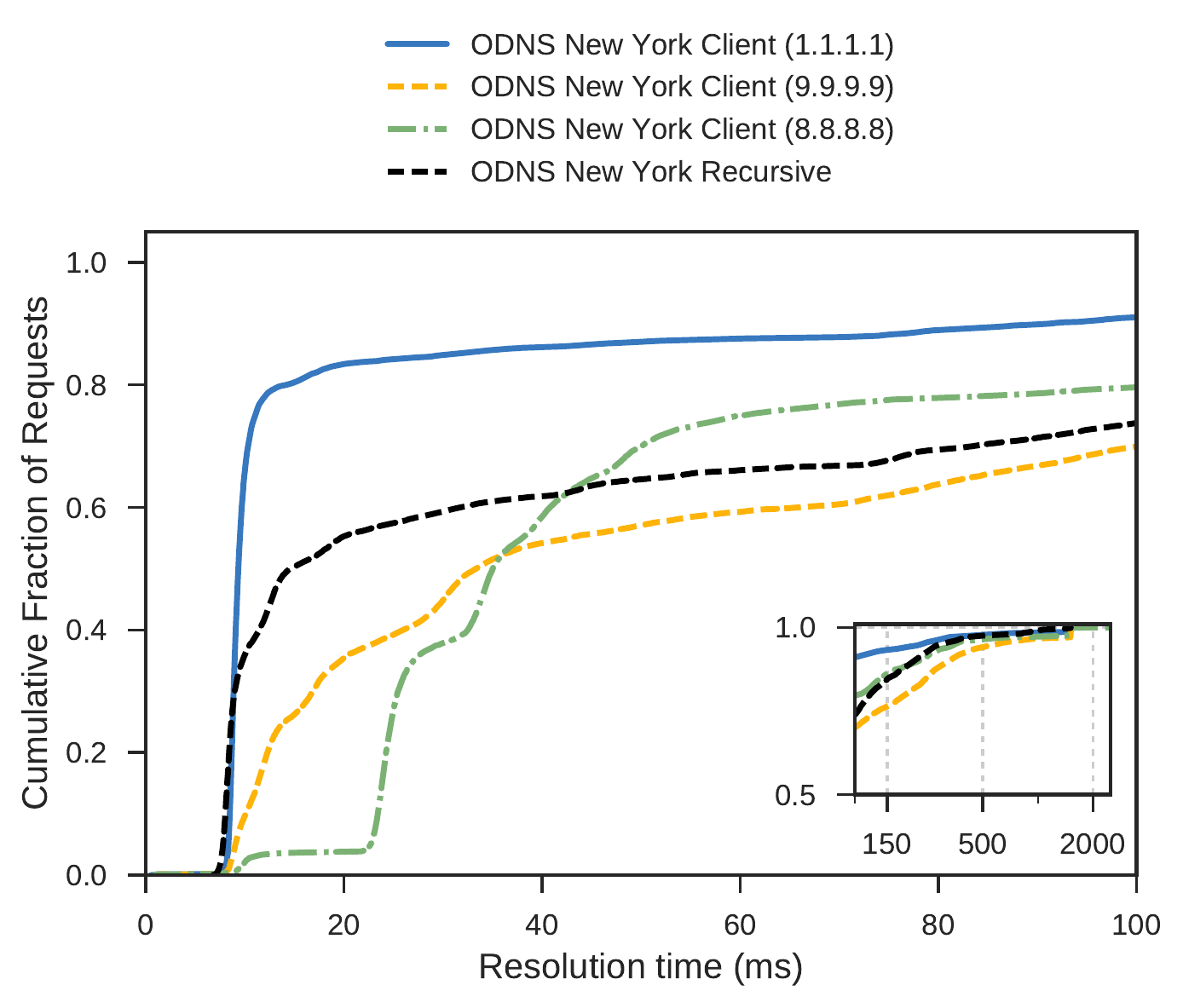}
\caption{Effects of caching at the ODNS resolver vs. forwarding all queries to a third-party open resolver.}
\label{fig:caching}
\end{figure}

\system{} allows for caching at both the stub resolver and at the \system{} resolver. To evaluate the effect of caching at the \system{} resolver on performance, we measure the overhead of \system{} where the resolver caches and acts as a second recursive resolver, and compare it to the overhead of \system{} simply forwarding queries to third party upstream recursive resolvers and their associated caches. We warm the cache using a DNS query trace generated for usage with {\tt resperf}. Figure~\ref{fig:caching} shows the results for the \system{} resolver in New York. We see that the resolver acting as a stub with Cloudflare's {\tt 1.1.1.1} service as its recursive resolver achieves the best results as they appear to have cache hits for roughly 80\% of requests and the additional latency to query {\tt 1.1.1.1} does not impact the overall performance. Interestingly, we find that \system{} acting as a recursive with a warmed cache outperforms both Quad9 and Google overall. These results lead us to conclude that the \system{} resolver can perform well as either a recursive resolver itself, or as a forwarder relying on third-party recursive resolvers, provided open resolvers perform well in the location that the \system{} resolver resides. In cases where disk storage is costly, the \system{} resolver can use a high-performance third-party resolver such as Cloudflare and gain the benefits of caching without incurring a large latency penalty. If the choice is made to use a third-party service, care must be taken to ensure that the same third-party service is not used by both the \system{} stub and the \system{} resolver, as such behavior could allow the third-party service to correlate timing between encrypted queries and subsequent plaintext queries from the \system{} resolver.  

\section{Related Work}
\label{sec:related}

In this section, we describe related work on DNS privacy. We discuss various infrastructure for protecting user privacy in DNS, ranging from alternate open resolvers to virtual private networks. In general, these systems involve simply changing the trusted third party who resolves DNS on behalf of the client, rather than decoupling the DNS query from the IP address that issued the query.

\textbf{Alternate DNS resolvers.} Quad9 provides both security and privacy features for DNS. Quad9 uses IBM X-Force threat intelligence data at the recursive resolver to prevent a client from accessing a malicious site~\cite{quad9}. Although this recursive resolver does not store or distribute the DNS data passing through, it still allows a DNS operator to observe this data. Once such information is retained, of course, it may become vulnerable to other threats to user privacy, including data requests from law enforcement. Cloudflare recently released {\tt 1.1.1.1}, which is a privacy-focused consumer DNS recursive resolver; it supports both DNS-over-TLS and DNS-over-HTTPS, and also offers the feature of query name minimization~\cite{cloudflare_dns}. While {\tt 1.1.1.1} only logs data for 24 hours at the recursive resolver (for debug purposes), it is still susceptible to a malicious DNS operator (or other adversary) saving that information before it is purged. Furthermore, even though a DNS operator who aims to protect user privacy may purge this information periodically, a user has no guarantee that information that an operator learns might be retained, for operational or other purposes.

\textbf{Privacy-preserving DNS.} EncDNS is functionally similar to \system{}~\cite{herrmann2014encdns}; however, EncDNS is different in several ways. First, EncDNS doesn't address the critical problem of key distribution. ODNS incorporates a key distribution mechanism directly in the system rather than using an out-of-band protocol like DANE (as suggested in~\cite{herrmann2014encdns}), which isn't widely deployed. Second, in this regard and others, ODNS is immediately deployable and makes no assumptions about the (future) deployment of protocols. Third, our performance evaluations (1) reflect a deployment on today's Internet infrastructure and (2) include holistic, end-to-end performance (e.g., page load times). In contrast to \system{} and EncDNS, most proposed DNS privacy mechanisms are protecting against an adversary, but not a DNS operator. Castillo-Perez and Garcia-Alfaro evaluate privacy-preserving DNS mechanisms, but show that they need additional measures to enhance their security~\cite{castillo2009evaluation}.  Similarly, Query Name Minimization is a proposal that limits what name servers see in DNS queries, but a recursive resolver's operator still learns the domain requested and the corresponding client who requested the domain~\cite{RFC7816}. Researchers have also pointed out how aspects of current (operational) DNS, such as prefetching, have privacy implications~\cite{krishnan2010dns,shulman2014pretty}. Federrath et al. introduced a DNS anonymity service that employs broadcasting popular hostnames and low-latency mixes for requesting less popular domains; unlike \system{}, this proposed DNS anonymity service is a clean-slate architecture and requires fundamental changes to the DNS infrastructure~\cite{federrath2011privacy}.

\textbf{DNS security.} Protocols to secure DNS include DNS-over-TLS~\cite{dns_tls}, DNSSEC~\cite{larson2005dns}, and T-DNS~\cite{zhu2015connection}. DNS-over-TLS protects the privacy of the domain being requested in transit, but does not prevent a recursive resolver's operator from learning both the client who issued the request and the content of the request. Some work has analyzed DNSSEC in more detail; Osterweil et al. develop SecSpider to monitor and detect errors in the DNSSEC deployment~\cite{osterweil2009deploying}. While there have been many attacks on DNS, the adoption of DNS security protocols is very limited; Herzberg and Shulman highlight some of the issues with retrofitting security into DNS~\cite{herzberg2014retrofitting}.  Researchers have also combined DNSSEC features with BIND DNS software to implement a system that prioritizes the integrity and availability of DNS~\cite{jalalzai2015dns}. Recent research has also seen the introduction of new frameworks for monitoring DNS in the hopes of detecting attacks~\cite{6212019,hesselman2017increasing}.

\textbf{Virtual private networks.} Systems to protect user privacy include Virtual Private Networks (VPNs). Unfortunately, many VPN providers send unencrypted DNS queries to the client's ISP, and all VPN providers would be able to associte DNS queries and responses with an individual client IP address. VPNGate uses different VPNs to contol where a client {\it appears} to be located~\cite{nobori2014vpn}, but even this approach does not prevent operators of DNS recursive resolvers from learning the domains being requested. The VPN operators themselves also still have complete information about the domains and the IP addresses that are querying them.

\section{Discussion}
\label{sec:discussion}

\textbf{Limitations.} \system{} prevents an adversary as described in Section \ref{sec:attacker}, but is limited in its capability of preventing other types of attacks.  For example, an adversary will be able to associate client IP addresses and their DNS queries if the adversary colludes between the recursive resolver and the \system{} resolver (or if the same adversary requests data from the recursive resolver and the \system{} resolver).  Additionally, if the recursive resolver supplies EDNS0 information to the \system{} resolver, then the \system{} resolver will be able to associate client IP addresses and their corresponding DNS queries.  Lastly, \system{} assumes a {\it passive} adversary, but real-world adversaries may also perform active attacks to compromise the integrity or availability of the system.  In the future, stronger adversaries will pose threats to \system{}, including collusion between the resolver and the \system{} resolver, as well as active attacks, although it's worth noting that in both of these cases, the attacks can be easily detected and in some cases the user can take steps to avoid them, even within the context of the current design.

\textbf{Integration with other security protocols.} Various DNS standards and enhancements have improved the security and privacy of DNS, including DNSSEC, DNS-over-TLS (DoT), and DNS-over-HTTPS (DoH). ODNS is compatible with all of these enhancements, and it provides complementary privacy guarantees. DNSSEC guarantees the integrity of records but does not provide privacy for DNS queries or clients; and DoT and DoH secure the transport between the client and the resolver, but do not hide the identity of the client or its query from the resolver. ODNS additionally protects the privacy of a client's DNS queries from the resolver that it chooses to use, and can also benefit from the guarantees that DoT, DoH, and DNSSEC provide. 

\textbf{Striping queries across multiple recursives.} The client stub resolver typically forwards DNS queries on to the client's recursive resolver, but \system{} supports forwarding the DNS queries on to any resolvers. \system{} can stripe DNS queries across the many available open resolvers, which helps increase the privacy of the client because  the recursive resolver does not see all ({\it obfuscated}) DNS queries from the client.  If striping is enabled, then each open resolver only sees some portion of each client's obfuscated queries. 

\textbf{Policy-based traffic routing.} There may be scenarios where \system{} users would prefer to control the geographic location where their \system{} resolver resides rather than for the selection to be based on anycast routing. For instance, a user may wish to avoid a resolver in a country that they suspect may eavesdrop on their traffic. \system{} can be easily extended to support such policy-based routing. To accomplish this, the \system{} resolvers can include location codes in their names (\eg, \texttt{us.odns} or \texttt{uk.odns}). Users could configure their stubs to select \system{} resolvers located in specific locations. However, such usage would come at a performance cost as the specified resolver may not result in the lowest latency. This scenario is akin to using a VPN to ingress and egress onto the Internet from a trusted location. We leave the implementation of this functionality for future work.

\textbf{Session key re-use.} In our current design, the \system{} stub generates unique session keys for each DNS query. This design impacts client performance in two ways. First, as discussed in Section~\ref{sec:micro}, the cryptographic operations result in additional latency for each query totaling up to roughly 1 ms at the stub. Additionally, the use of unique session keys for each query renders caching, even for identical queries from a single user, at the recursive resolver impossible. To ease these challenges, session keys could be re-used for some period of time by the stub. In cases where keys are reused, the potential for cache hits at the recursive resolver exists for multiple identical queries by clients with the same combination of symmetric key and public key of the \system{} resolver. However, excessive key re-use may leave the keys vulnerable to attack. Thus, cost / benefit analysis pertaining to key reuse must be done prior to implementing the feature. We leave this for future exploration.  
 
\textbf{Denial of service attacks.} \system{}'s resolvers cannot check the incoming IP address of queries, which could facilitate Denial of Service attacks. To defend against DoS attacks, the client's stub resolver can append bytes that indicate the DNS query is sent from an \system{}-participating client stub resolver. The \system{} resolver can then check for these bytes and verify that it was sent via \system{} prior to decryption. This functionality is left to future work.
\section{Conclusion}
\label{sec:conclusion}
DNS queries can reveal personal information such as browsing patterns as well as types of devices and allow DNS resolvers to associate such information with client IP addresses. Users are required to place trust in their recursive DNS resolver. Privacy-focused, third party DNS resolvers simply shift the trust without alleviating the fundamental information exposure. In this work, we present \system{}, a system that decouples client IP address from DNS queries, removing the need for trust altogether as no DNS infrastructure outside of the user network is able to obtain both pieces of information. \system{} is designed to be fully-compatible with existing DNS infrastructure and requires only minimal changes. Our evaluation of \system{} reveals that latency overhead is minimal, performance for user web traffic is acceptable, and minimal impact on recursive resolver traffic.  

We are seeking to move forward with this work to deploy in-the-wild. In the near term, we have presented an Internet-Draft on ODNS at the IRTF and IETF to be considered for adoption into the DNS Privacy (DPRIVE) working group and are implementing its functionality into commercial DNS server code bases. Long-term, we are exploring widespread deployment of \system{} resolvers with the cooperation of an operator of an existing global anycast network.

\section{Acknowledgements}
We thank our shepherd, Daniel Zappala, and the anonymous reviewers for the feedback and comments. This research was supported in part by the Center for Information Technology Policy at Princeton University and by NSF award CNS-1535796. 
\clearpage

\bibliographystyle{acm}
\bibliography{paper}

\end{document}